\documentclass[10pt,preprintnumbers,aps,amssymb,nofootinbib,amsmath,superscriptaddress,notitlepage]
{revtex4-2}
\usepackage{epsfig,epsf}
\usepackage{comment}
\usepackage{bm} 
\usepackage{color} 
\usepackage{slashed}
\usepackage{relsize}	
\usepackage{soul} 
\usepackage{hyperref}
\usepackage{tensor} 
\usepackage{yfonts} 
\newcommand{\beq}{\begin{equation}}
\newcommand{\beql}[1]{\begin{equation}\label{#1}}
\newcommand{\eeq}{\end{equation}}
\def\bal#1\gal{\begin{align}#1\end{align}}
\newcommand{\ball}[1]{\bal\label{#1}}
\newcommand{\eq}[1]{(\ref{#1})}
\newcommand{\fig}[1]{Fig.~\ref{#1}}
\renewcommand{\sec}[1]{Sec.~\ref{#1}}
\DeclareMathOperator{\sgn}{sgn}

\DeclareMathOperator{\real}{\mathrm{Re}}

\renewcommand{\b}[1]{{\bm #1}} 
\newcommand{\unit}[1]{\hat {{\bm #1}}} 

\begin{document}

\title{Polarized electromagnetic radiation by chiral media with time-dependent chiral chemical potential}

\author{Kirill Tuchin}

\affiliation{
Department of Physics and Astronomy, Iowa State University, Ames, Iowa, 50011, USA}

\date{\today}

\begin{abstract}

We investigate photon production from media characterized by a time-dependent chiral chemical potential $\mu_5(t)$. 
We first consider the chiral anomaly as a small perturbation and show that $\mu_5\to 2\gamma$ process generates non-polarized radiation, as the probabilities of producing right and left-handed  photons are equal. In chiral semimetal with a vanishing chiral chemical potential but  a finite  separation of Weyl nodes, $\b \Delta(t)$, the photon production vanishes at the leading order of perturbation theory. We then employ the method of Bogolyubov transformation to obtain the photon spectrum that sums up all powers of $\mu_5$.  Employing the exactly solvable model $\mu_5(t) \propto A+B\tanh(t/\tau)$, we demonstrate that the spectrum exhibits strong circular polarization, whose direction is determined by the sign of $\mu_5$. The spectrum contains resonances associated with the instability of the electromagnetic field in chiral media. We discuss potential phenomenological applications of these findings. 

\end{abstract}

\maketitle

\section{Introduction}\label{sec:a}

The chiral anomaly of QED \cite{Adler:1969gk,Bell:1969ts} couples the pseudoscalar current to the field invariant $F\tilde F$, where $F$ is the electromagnetic field tensor and  $\tilde F$ its dual. This anomaly is famous for allowing the two-photon decay of the neutral pion. In a chiral medium, the anomaly is responsible for inducing the chiral magnetic and anomalous Hall effects. The chiral magnetic effect refers to the generation of electric current in response to the magnetic field: $\b j = c_A\mu_5\b B$, where $\mu_5$ is the chiral chemical potential, the transport coefficient $b_0=c_A\mu_5$ known as the chiral magnetic conductivity, and $c_A$ is the anomaly coefficient. The anomalous Hall current emerges as a response to the external electric field  $\b j = \b b\times \b E$. In Weyl semimetals, the transport coefficient is $\b b=(\alpha/\pi)\b \Delta$, where  $\b\Delta$ represents the displacement of the Weyl nodes in the momentum space $\b\Delta$.\footnote{A typical Weyl semimetal posseses multiple pairs of  nodes, in which case $\b\Delta$ represents  the total displacement  \cite{Zubkov:2023dpe}. It is finite in magnetic Weyl semimetals but vanishes in non-magnetic ones.} Quite similar to the neutral pion's two-photon decay, the time-dependent $\mu_5$ induces production of photon pairs in the chiral medium. At the leading order in $\mu_5$, the perturbation theory predicts that the decay $\mu_5\to 2\gamma$ produces right and left-handed photons with equal probability. Consequently, the photon spectrum is not polarized. However, the leading order perturbation theory does not provide the complete picture. It is adequate to compute the neutral pion decay because its coupling is small, but it is not sufficient when $\mu_5$ or $\b\Delta$ are of the same magnitude as the photon's momentum.

The primary objective  of this paper is to exactly compute the photon spectrum in a time-dependent $\mu_5$ and demonstrate its strong circular polarization. We use the method of the Bogolyubov transformation, which entails the canonical quantization of the electromagnetic field at different equal-time hypersurfaces. Since time-dependence of the normal modes is generally non-harmonic, the Fock spaces, including their vacuum states, vary across  different hypersurfaces.  As a result, the vacuum state at the remote future $t\to \infty$ may contain photons, even though it did not at the initial time $t\to -\infty$. Computing the exact spectrum requires solving the wave equation at finite $\mu_5$. Analytical solutions are known only for a limited number of special cases, one of which we employ in this paper to gain insights into the photon production dynamics: $b_0= A+B\tanh(t/\tau)$, where $A$, $B$ and $\tau$ are real numbers. Our main result is the photon spectrum \eq{d25} which demonstrates complicated nonlinear dependence on the parameters $A$ and $B$. This spectrum is presented  in Figs.~\ref{fig0} and \ref{fig1}, which clearly indicate strong circular polarization of the emitted radiation.

The paper is structured as follows. In \sec{sec:p} we compute photon production in $\mu_5\to 2\gamma$ and  $\b\Delta\to 2\gamma$ using the perturbative approach and show that the latter vanishes. In \sec{sec:b} we review the method of the Bogolyubov transformation. In \sec{sec:f} we consider a specific model for  $b_0(t)$, that can be solved exactly and yields the photon spectrum \eq{d25} presented in \fig{fig0} and \fig{fig1}. The discussion and summary are presented in \sec{sec:s}.

The production of photons in the presence of the chiral anomaly was previously discussed in the leading order of perturbation theory. Refs.~\cite{Fukushima:2012fg,Yee:2013qma,Jia:2022awu,Jia:2024ctx,Wang:2024gnh} computed photon emission assisted by a magnetic field, while Refs.~\cite{Kharzeev:2013wra,Tuchin:2019jxd} considered radiation by charged fermions.  The polarization of the  photon spectrum was emphasized in \cite{Tuchin:2019jxd,Hansen:2024kvc,Hansen:2025gzt}.

\section{Perturbation theory}\label{sec:p}

 A convenient systematic way for incorporating anomalous currents into Maxwell's theory is by means of  the  Lagrangian \cite{Wilczek:1987mv,Carroll:1989vb, Sikivie:1984yz}:
\ball{p3}
\mathcal{L}= -\frac{1}{4} \left(F^2+\theta F \tilde F\right)\,,
\gal
where $F$ is the electromagnetic field tensor, $\theta$ is an external pseudo-scalar field such that $\partial_\mu\theta = b_\mu=(b_0,-\b b)$. We assume that $b_\mu$ is a function of time, but not of spatial coordinates.  All medium effects other than the anomalous chiral magnetic and Hall currents are omitted for simplicity.  Using the identity $F\tilde F = 2\partial_\mu(\varepsilon^{\mu\nu\lambda\rho}A_\nu\partial_\lambda A_\rho)$ and integrating the second term by parts one can cast \eq{p3} in the form:
\ball{p5}
\mathcal{L}=-\frac{1}{4} F^2+\mathcal{L}_\text{int}=  -\frac{1}{4} F^2+ \frac{1}{2}  \varepsilon^{\mu\nu\lambda\rho}A_\nu\partial_\lambda A_\rho \partial_\mu\theta\,.
\gal

The essence of the perturbation theory lies  in treating the second term in  \eq{p5} as a small perturbation of the first term, describing the free electromagnetic field. The canonical quantization of the field is accomplished by expanding it in the complete set of normal modes of the free Lagrangian:
\bal\label{p7}
\b A(t,\b x)= \int \frac{d^3k V}{(2\pi)^3}\sum_\lambda\left( a_{\b k,\lambda}\b A_{\b k,\lambda}(t,\b x)+ a^\dagger_{\b k,\lambda}\b A^*_{\b k,\lambda}(t,\b x)\right)\,,
\gal
where $\b k$ and $\lambda$ is photon's momentum and helicity respectively, and $V$ is the volume. The normal modes of given momentum are plane waves. In the radiation gauge $A^0=0$, $\b \nabla\cdot \b A=0$ they are:
\ball{p9}
\b A_{\b k,\lambda}(t, \b x)= \frac{1}{\sqrt{2\omega V}}\b \epsilon_{\b k,\lambda} e^{ i\b k\cdot \b x-i\omega t}\,,
\gal
where $\b\epsilon_{\b k,\lambda} $ is the polarization vector, and $\omega=k$. The coefficients of the expansion \eq{p7} are the creation and annihilation operators subject to the usual commutation relations. There still remains a residual gauge freedom, which manifests in the ability to choose any pair of  polarization vectors that meet the conditions $\b\epsilon_{\b k,\lambda} \cdot \b k=0$ and $|\b\epsilon_{\b k,\lambda} |=1$. Note, that in this approach, the dispersion relation is independent of the photon polarization. 

The production of two photons with four momenta $k= (\omega,\b k)$ and $k'=(\omega',\b k')$ and polarizations $\epsilon=(0,\b \epsilon_{\b k,\lambda})$ and  $\epsilon'=(0,\b \epsilon_{\b k',\lambda'})$ respectively is represented  by the matrix element
\bal
S(b\to 2\gamma)&= -\frac{i}{2}\varepsilon^{\mu\nu\lambda\rho}\frac{\epsilon_\nu}{\sqrt{2\omega V}}(k'-k)_\lambda \frac{\epsilon_\rho'}{\sqrt{2\omega' V}}(2\pi)^3\delta(\b k+\b k')\tilde b_\mu(\omega+\omega')\label{p11}\\
&=-\frac{i}{2}\frac{1}{\sqrt{2\omega V}}\frac{1}{\sqrt{2\omega' V}}(2\pi)^3\delta(\b k+\b k')
\left[ \tilde b_0(\omega+\omega') (\b k-\b k')+\tilde {\b b}(\omega+\omega')(\omega-\omega')\right]\cdot \left( \b\epsilon_{\b k,\lambda}\times \b \epsilon_{\b k',\lambda'}\right)\,,\label{p12}
\gal
where 
\ball{p15}
\tilde b_\mu(\omega+\omega')= \int_{-\infty}^{+\infty} e^{-i(\omega+\omega')t}b_\mu(t)dt\,.
\gal
Since $\b k = -\b k'$, $\omega= \omega'$, the term proportional to $\b b$ vanishes. Using the circular polarization basis in which $\b\epsilon_{\b k,\pm}^*= \b\epsilon_{\b k,\mp}=\b\epsilon_{-\b k,\pm}$ we compute
\ball{p17}
\b\epsilon_{\b k,\lambda}\times \b \epsilon_{\b k',\lambda'}= -\frac{i}{2}\unit k(\lambda'+\lambda)= -i\unit k \delta_{\lambda,\lambda'}\,.
\gal
The decay probability into photons of helicities $\lambda$ and $\lambda'$ is:
\bal
dw(b\to 2\gamma)&= |S(b\to 2\gamma)|^2\frac{d^3k V}{(2\pi)^3}\frac{d^3k' V}{(2\pi)^3}= \frac{1}{4}|\tilde b_0|^2\delta_{\lambda,\lambda'}\frac{d^3k V}{(2\pi)^3}\,.\label{p20}
\gal
The decay probabilities into two positive-helicity photons $\lambda=\lambda'=1$ or two negative helicity photons $\lambda=\lambda'=-1$ are equal. In either case, one photon is right-hand polarized and another one is left-hand polarized because their momenta are back-to-back. Consequently, the average photon polarization vanishes. The total number of right or left-handed photons per phase space reads:
\bal\label{p22}
\frac{d\mathcal{N}_\lambda}{Vd^3k} &=\frac{1}{4}|\tilde b_0|^2\frac{d^3k V}{(2\pi)^3}\,.
\gal

For future reference and as an illustration consider the following time-evolution:
\bal\label{p25}
b_0(t)=A+ B\tanh\frac{t}{\tau}\,.
\gal
This model has recently  been used to investigate the chiral Cherenkov radiation in time-dependent $\mu_5$ \cite{Tuchin:2025lbz}. The corresponding Fourier transform \eq{p15} of \eq{p25} can be expressed as:
\ball{p27}
\tilde b_0(\omega+\omega')&=\lim_{\epsilon\to 0}\left\{  \int_{-\infty}^0 e^{-i(\omega+\omega')t+\epsilon t}b_0(t)dt + \int_0^\infty e^{-i(\omega+\omega')t-\epsilon t}b_0(t)dt \right\} = -\frac{i\pi \tau B}{\sinh(\pi\tau \omega)}\,.
\gal
Therefore,
\ball{p30}
d\mathcal{N}_{\lambda}&= \frac{1}{4}\frac{(\pi \tau B)^2}{\sinh^2(\pi\tau \omega)}\frac{d^3k V}{(2\pi)^3}\,.
\gal

In the subsequent sections, we adopt an alternative approach that allows the summation of higher order powers of $b_0$. It will lead us to a conclusion that the production probability of the two circular polarizations is in fact not equal, implying that the radiation exhibits polarization.

\section{Bogolyubov transformation at finite $\mu_5(t)$}\label{sec:b}

The complete set of normal modes \eq{p9} used by the perturbative method corresponds to the free waves described by the first term in \eq{p5}. To obtain the exact solution, we  seek the complete set of normal modes associated with  \emph{both} terms in \eq{p5}.  We consider an isotropic chiral medium with $\b \Delta=0$ and neglect all electromagnetic responses other than the chiral magnetic effect for the sake of simplicity. The electromagnetic field is described by the equations derived from \eq{p5}:
\bal
&\b \nabla \cdot \b E = 0\,,\label{a1}\\
&\b \nabla\times \b B- \partial_t \b E = b_0\b B\,,\label{a2}\\
&\b \nabla \times \b E+ \partial_t \b B=0\,,\label{a3}\\
&\b  \nabla \cdot \b B=0\,. \label{a4}
\gal
 In the radiation gauge  the vector potential satisfies the equation 
\ball{b1}
\nabla^2 \b A -\partial_t^2\b A+ b_0(t)\b \nabla\times \b A=0\,.
\gal
We are interested in the normal modes with given momentum $\b k$ and polarization vector $\b\epsilon_\lambda$. Owing to the identity $i\unit k \times \b\epsilon_\lambda = \lambda \b\epsilon_\lambda$, the two independent polarizations are the circular right and left-handed ones corresponding to helicities $\lambda=\pm 1$. Therefore, the general form of the normal modes is
\ball{b3}
\b A_{\b k,\lambda}(t, \b x)= \frac{1}{\sqrt{V}}a(t)\b \epsilon_\lambda e^{ i\b k\cdot \b x}\,,
\gal
where the amplitude $a(t)$ is governed by the equation
\ball{b5}
\ddot a(t)+\Omega^2(t)a(t) =0\,,
\gal
with
\ball{b7}
\Omega^2(t)= k^2 -\lambda b_0(t)k\,.
\gal
The photon wave function will be normalized to one photon per unit volume as usual. 

It is important to note that the choice of the radiation gauge has completely exhausted the gauge freedom.  This is because photons must be circularly polarized due to the presence of the curl term in \eq{b1}. In contrast, polarization of photons in the perturbation theory is not completely fixed by the radiation gauge, allowing us to choose any two independent photon polarizations in the plane perpendicular to $\b k$.

Suppose that at distant past and future the function $b_0(t)$  is constant: $b_0(-\infty) = b_0^\text{in}$, $b_0(\infty) = b_0^\text{out}$. Then, it follows from \eq{b5} that the positive frequency normal modes at $t\to -\infty$ have the following asymptotic behavior in the distant past:
\ball{b10}
\lim_{t\to -\infty} \b A^\text{in}_{\b k,\lambda}(t,\b x)=  \frac{1}{\sqrt{2\omega^\text{in} V}}\b \epsilon_{\b k,\lambda} e^{-i\omega^\text{in} t+ i\b k\cdot \b x}\,,
\gal 
where 
\ball{b11}
\omega^\text{in}= \sqrt{ k^2 -\lambda b_0^\text{in} k}\,.
\gal
The operator of the electromagnetic field can be expanded in these `in'-modes:
\bal\label{b13}
\b A(t,\b x)= \int \frac{d^3k V}{(2\pi)^3}\sum_\lambda\left( a^\text{in}_{\b k,\lambda}\b A^\text{in}_{\b k,\lambda}(t,\b x)+ a^{\text{in}\dagger}_{\b k,\lambda}\b A^{\text{in}*}_{\b k,\lambda}(t,\b x)\right)\,.
\gal
In the vacuum state, defined as  $a^\text{in}_{\b k,\lambda}|0\rangle^\text{in}=0$,  the expectation value of the particle number operator $N_{\b k,\lambda}=a^{\text{in}\dagger}_{\b k,\lambda}a^\text{in}_{\b k,\lambda}$ vanishes:
\bal\label{b15}
^\text{in}\langle 0|N_{\b k,\lambda}|0\rangle^\text{in}=0\,.
\gal
In the remote future, the positive frequency normal modes are different, and have the following asymptotic behavior:
\ball{b17}
\lim_{t\to \infty} \b A^\text{out}_{\b k,\lambda}(t,\b x)= \frac{1}{\sqrt{2\omega^\text{out} V}}\b \epsilon_{\b k,\lambda} e^{-i\omega^\text{out} t+ i\b k\cdot \b x}\,,
\gal 
where 
\ball{b19}
\omega^\text{out}= \sqrt{ k^2 -\lambda b_0^\text{out}k}\,.
\gal
The expansion of the operator of electromagnetic field in terms of `out'-modes reads:
\bal\label{b21}
\b A(t,\b x)= \int \frac{d^3k V}{(2\pi)^3}\sum_\lambda\left( a^\text{out}_{\b k,\lambda}\b A^\text{out}_{\b k,\lambda}(t,\b x)+ a^{\text{out}\dagger}_{\b k,\lambda}\b A^{\text{out}*}_{\b k,\lambda}(t,\b x)\right)\,.
\gal
The corresponding vacuum state is defined as  $a^\text{out}_{\b k,\lambda}|0\rangle^\text{out}=0$.

The normal modes are normalized such that their scalar products are:
%
%
\bal
&\left(\b A^\text{in}_{\b k,\lambda},\b A^\text{in}_{\b k',\lambda'}\right)= i\int d^3x \b A^{\text{in}*}_{\b k,\lambda}\stackrel{\leftrightarrow}{\partial_t}\! \b A^\text{in}_{\b k',\lambda'}
=\frac{(2\pi)^3}{V}\delta_{\lambda \lambda'}\delta(\b k-\b k')\,,\label{b23}\\
&\left(\b A^\text{in}_{\b k,\lambda},\b A^{*\text{in}}_{\b k',\lambda'}\right)=0\,,\label{b24}\\
&\left(\b A^{\text{in}*}_{\b k,\lambda},\b A^{\text{in}*}_{\b k',\lambda'}\right)
=-\frac{(2\pi)^3}{V}\delta_{\lambda \lambda'}\delta(\b k-\b k')\,,\label{b25}
\gal
 The corresponding commutation relations of the creation and annihilation operators of the `in'-modes  are:
\bal
\left[a^\text{in}_{\b k,\lambda},a^{\text{in}\dagger}_{\b k',\lambda'}\right]=\frac{(2\pi)^3}{V}\delta(\b k-\b k')\delta_{\lambda\lambda'}\,,\label{b27}\\
\left[a^{\text{in}\dagger}_{\b k,\lambda},a^{\text{in}\dagger}_{\b k',\lambda'}\right]=
\left[a^{\text{in}}_{\b k,\lambda},a^\text{in}_{\b k',\lambda'}\right]=0\,,\label{b28}
\gal
and, similarly, for the `out'-modes.
The relationship between the two sets of modes is given by the Bogolyubov transformation:
\bal
\b A^\text{out}_{\b k,\lambda}&= \sum_{\lambda'}\int \frac{d^3k' V}{(2\pi)^3}\left( \alpha_{\b k,\lambda; \b k',\lambda'}\b A^\text{in}_{\b k',\lambda'}+\beta_{\b k,\lambda; \b k',\lambda'}\b A^{\text{in}*}_{\b k',\lambda'}\right)\,\label{b31}\\
\b A^\text{in}_{\b k',\lambda'}&= \sum_\lambda\int \frac{d^3k V}{(2\pi)^3}\left( \alpha^*_{\b k,\lambda; \b k',\lambda'}\b A^\text{out}_{\b k,\lambda}-\beta_{\b k,\lambda; \b k',\lambda'}\b A^{\text{out}*}_{\b k,\lambda}\right)\,.\label{b32}
\gal
The Bogolyubov coefficients can be computed as follows:
\bal
 &\alpha_{\b k,\lambda; \b k',\lambda'} = \left(\b A^\text{out}_{\b k,\lambda},\b A^\text{in}_{\b k',\lambda'}\right)\,, \label{b35}\\
  &\beta_{\b k,\lambda; \b k',\lambda'} = -\left(\b A^\text{out}_{\b k,\lambda},\b A^{\text{in}*}_{\b k',\lambda'}\right)\,. \label{b36}
\gal
Using the relations \eq{b13},\eq{b21}--\eq{b32} one can express the creation and annihilation operators as
\bal
a^\text{in}_{\b k,\lambda}&= \sum_{\lambda'} \int \frac{d^3k' V}{(2\pi)^3}  \left(\alpha_{\b k,\lambda; \b k',\lambda'}a^\text{out}_{\b k',\lambda'}
+\beta_{\b k',\lambda'; \b k,\lambda} ^*a^{\text{out}\dagger}_{\b k',\lambda'}\right)\,,\label{b38}\\
a^\text{out}_{\b k',\lambda'}&= \sum_\lambda \int \frac{d^3k V}{(2\pi)^3} \left(\alpha_{\b k',\lambda'; \b k,\lambda}a^\text{in}_{\b k,\lambda}
-\beta_{\b k',\lambda'; \b k,\lambda} ^*a^{\text{in}\dagger}_{\b k,\lambda}\right)\,. \label{b39}
\gal
It follows that the future vacuum contains photons:
\bal\label{b42}
^\text{out}\langle 0|N_{\b k,\lambda}|0\rangle^\text{out}= {^\text{out}}\langle 0|a^{\text{in}\dagger}_{\b k,\lambda}a^\text{in}_{\b k,\lambda}|0\rangle^\text{out}=\sum_{\lambda'}\int \frac{d^3k' V}{(2\pi)^3}  \left|\beta_{\b k,\lambda; \b k',\lambda'}\right|^2\,.
\gal
This general result shows that if $b_0$ changes in time, the number \eq{b42} of photons are produced.

\section{Photon spectrum for $b_0(t)=A+ B\tanh\frac{t}{\tau}$}\label{sec:f}

The exact solutions of \eq{b5} relevant in practical applications are few. In this section, we study one such analytically solvable model for the time-evolution of the chiral magnetic conductivity given by \eq{p25}. The initial and final values of $b_0(t)$ are  $b_0^\text{in}=A-B$ and $b_0^\text{out}= A+B$ respectively. The transition time between them is controlled by the parameter $\tau$. Substituting \eq{p25} into \eq{b7} we obtain the function $\Omega(t)$:
\ball{d3}
\Omega^2(t)= k^2-\lambda k A-\lambda k B \tanh \frac{t}{\tau}\,.
\gal
The exact solution of \eq{b5} with $\Omega$ given by \eq{d3} was obtained in \cite{Bernard:1977reg}. The solutions to \eq{b5} describing the normal modes having asymptotic behavior \eq{b10} and \eq{b17} read:
\bal
\b A^\text{in}_{\b k,\lambda}(t,\b x)&= \frac{1}{\sqrt{2\omega^\text{in}V}}\b \epsilon_\lambda e^{ i\b k\cdot \b x}e^{-i\omega^+ t-i\omega^-\tau \ln\left[2\cosh\frac{t}{\tau}\right]}
{_2}F_1\left(1+i\omega^-\tau, i\omega^-\tau;1+i\omega^\text{in}\tau;\frac{1}{2}\left(1+\tanh \frac{t}{\tau}\right)\right)\label{d9}\,,\\
\b A^\text{out}_{\b k,\lambda}(t,\b x)&= \frac{1}{\sqrt{2\omega^\text{out}V}}\b \epsilon_\lambda e^{ i\b k\cdot \b x}e^{-i\omega^+ t-i\omega^-\tau \ln\left[2\cosh\frac{t}{\tau}\right]}
{_2}F_1\left(1+i\omega^-\tau, i\omega^-\tau;1+i\omega^\text{out}\tau;\frac{1}{2}\left(1-\tanh \frac{t}{\tau}\right)\right)\,,\label{d10}
\gal
where $\omega^\text{in}$ and $\omega^\text{out}$ are given by  \eq{b11},\eq{b19} and we denoted 
\bal
\omega^\pm & = \frac{1}{2}(\omega^\text{out}\pm \omega^\text{in})\,. \label{d7}
\gal

The relationship between the two solutions \eq{d9} and \eq{d10} can be obtained using the properties of the hypergeometric functions:
\bal\label{d14}
\b A^\text{in}_{\b k,\lambda}(t,\b x)&=\alpha_{\b k,\lambda}\b A^\text{out}_{\b k,\lambda}(t,\b x)+\beta_{\b k,\lambda}\b A^{\text{out}*}_{-\b k,\lambda}(t,\b x)\,,
\gal
where 
\bal
\alpha_{\b k,\lambda}&=\left(\frac{\omega^\text{out}}{\omega^\text{in}}\right)^{1/2}\frac{\Gamma\left(1-i\omega^\text{in}\tau\right)\Gamma\left(-i\omega^\text{out}\tau\right)}{\Gamma\left(-i\omega^+\tau\right)\Gamma\left(1-i\omega^+\tau\right)}  \,,\label{d17}\\
\beta_{\b k,\lambda}&=\left(\frac{\omega^\text{out}}{\omega^\text{in}}\right)^{1/2}\frac{\Gamma\left(1-i\omega^\text{in}\tau\right)\Gamma\left(i\omega^\text{out}\tau\right)}{\Gamma\left(i\omega^-\tau\right)\Gamma\left(1+i\omega^-\tau\right)}\,. \label{d18}
\gal
It follows from \eq{b31} that the Bogolyubov coefficients read
\ball{d21}
\alpha_{\b k,\lambda; \b k',\lambda'}= \frac{(2\pi)^3}{V}\delta(\b k-\b k')\alpha_{\b k,\lambda}\,,\qquad
\beta_{\b k,\lambda; \b k',\lambda'}=\frac{(2\pi)^3}{V}\delta(\b k-\b k')\beta_{\b k,\lambda}\,.
\gal
Therefore, the number of photons with momentum $\b k$ and polarization $\lambda$ is given by \eq{b42}:
\ball{d23}
^\text{out}\langle 0|N_{\b k,\lambda}|0\rangle^\text{out}=\left| \beta_{\b k,\lambda}\right|^2\,,
\gal
Finally, since the photon spectrum is continuous, the experimental observable is the number of photons with momentum $\b k$ and polarization $\lambda$ in the unit phase space:
\ball{d25}  
\frac{d\mathcal{N}_\lambda}{Vd^3k}=\frac{ \left| \beta_{\b k,\lambda}\right|^2}{(2\pi)^3}= 
\frac{1}{(2\pi)^3}\frac{\omega^\text{out}}{\omega^\text{in}}\left| 
\frac{\Gamma\left(1-i\omega^\text{in}\tau\right)\Gamma\left(i\omega^\text{out}\tau\right)}{\Gamma\left(i\omega^-\tau\right)\Gamma\left(1+i\omega^-\tau\right)}
\right|^2\,.
\gal
The salient feature of the spectrum is its explicit dependence on the photon polarization through \eq{b11} and \eq{b19}. Thus the electromagnetic radiation is circularly polarized. 

The photon spectrum \eq{d25} exhibits a finite number of divergencies.  One of these divergencies arises when $\omega^\text{in}$ vanishes, which occurs at 
\ball{d31}
k^\text{in}= \lambda b_0^\text{in}\,,
\gal
and only affects photons with polarization $\lambda= \sgn b_0^\text{in}$. Additional divergences originate from the second gamma function in the numerator of \eq{d25}, which develops simple poles when its argument vanishes or equals a negative integer. As a result, the photon spectrum \eq{d25} diverges at the following values of the photon momentum:
\bal
&k^\text{out}_\pm= \frac{\lambda b_0^\text{out}}{2} \pm \sqrt{\left(\frac{b_0^\text{out}}{2}\right)^2-\frac{n^2}{\tau^2} }\,, \quad n=0,1,2,\ldots\,,\label{d33}
\gal
and again, only for photons with polarization $\lambda = \sgn b_0^\text{out}$. While  the pole at $k^\text{in}= \lambda b_0^\text{in}$  corresponding to $n=0$ always exists, those with positive $n$  appear only if the product $\lambda b_0^\text{out}\tau$ is sufficiently large.

In practice, photon states have finite width $\epsilon$ which can be accounted for by adding $i\epsilon$ to $\omega^\text{in}$ and $\omega^\text{out}$. As a result, the divergences appear as resonances at photon momenta \eq{d31} and \eq{d33}. The emergence of these resonances is a consequence of instability of electromagnetic field in chiral medium  that develops when $k<\lambda b_0^\text{in}$ or $k<\lambda b_0^\text{out}$ for the polarizations $\lambda = \sgn b_0^\text{in}$ or $\lambda = \sgn b_0^\text{out}$ respectively and has been extensively studied in the literature, see  \cite{Joyce:1997uy,Boyarsky:2011uy,Hirono:2015rla,Akamatsu:2013pjd,Tuchin:2014iua,Buividovich:2015jfa,Manuel:2015zpa,Kirilin:2017tdh} and references therein.

\begin{figure}[t]
\begin{tabular}{cc}
      \includegraphics[width=0.45\linewidth]{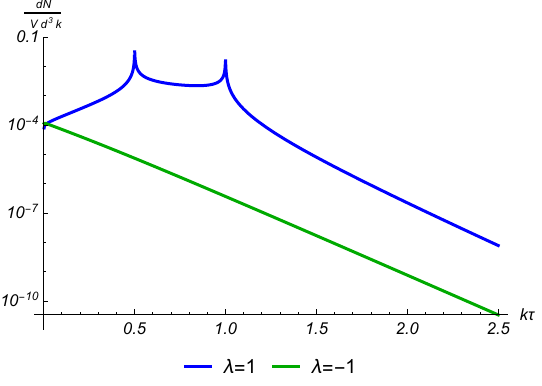} &
         \includegraphics[width=0.45\linewidth]{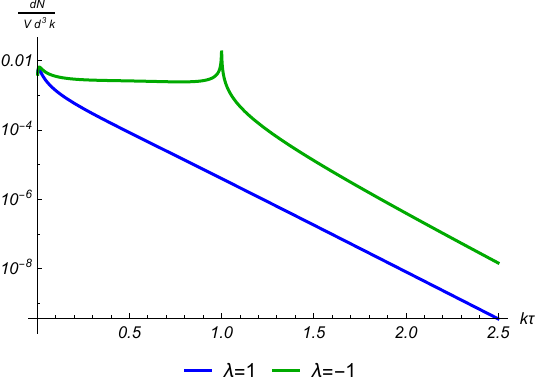}
   \end{tabular}       
  \caption{Photon spectrum radiated from a chiral medium. Left: $b_0^\text{in}\tau =1/2$, $b_0^\text{out}\tau =1$ , Right: $b_0^\text{in}\tau =0$, $b_0^\text{out}\tau =-1$. }
\label{fig0}
\end{figure}

\begin{figure}[t]
\begin{tabular}{cc}
      \includegraphics[width=0.45\linewidth]{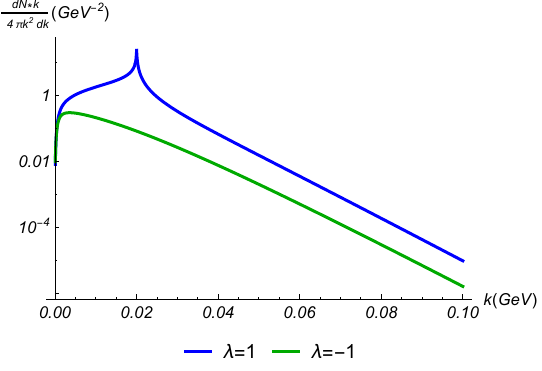} &
         \includegraphics[width=0.45\linewidth]{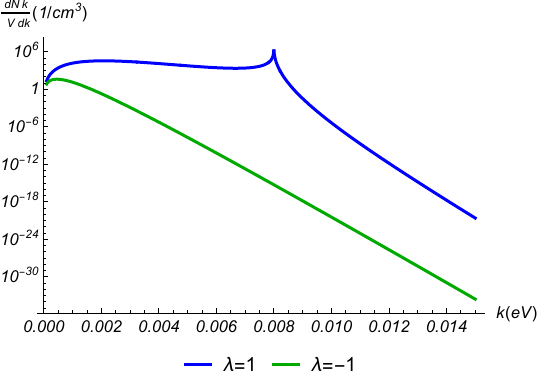}
   \end{tabular}       
  \caption{Left: photon spectrum radiated from quark-gluon plasma with $A=-B=10$~MeV,  $\tau=5$~fm, $V=(46\,\text{fm})^3$, $\epsilon=1$~MeV, $c_A=(2\alpha/\pi)N_c\sum_f q_f^2$. Right: photon spectrum radiated from a chiral semimetal with $A=-B= 4$~meV, $\tau=10^3$~eV$^{-1}=0.66$~ps, $\b \Delta=0$, $\epsilon=0.01$~meV, $c_A=2\alpha/\pi$.   }
\label{fig1}
\end{figure}

It is instructive to investigate the small chiral chemical potential limit of \eq{d25} and compare it to the perturbative calculations in \sec{sec:p}. At $|b_0|\ll k$, we can expand \eq{d18} to obtain 
\ball{d34}
\beta_{\b k,\lambda}\approx \frac{1}{4}\left(b_0^\text{in}-b_0^\text{out}\right)\frac{\pi \tau\lambda}{\sinh(\pi\tau k)}\left\{ 1+ 
\frac{\lambda}{2k} f(k\tau)
\right\} + \mathcal{O}(b_0^3)
\,.
\gal
We defined an auxiliary function:
\ball{d36}
f(k\tau)=
b_0^\text{out}+ib_0^\text{out} k\tau\left[\psi (1-ik\tau) -\gamma\right]+ib_0^\text{in}k\tau\left[\gamma- \psi(i k \tau)\right]\,,
\gal
where $\gamma$ is Euler's constant, and $\psi$ is the digamma function.  Substituting \eq{d34} into \eq{d25} yields:
\ball{d38}  
\frac{d\mathcal{N}_\lambda}{Vd^3k}\approx\frac{1}{(2\pi)^3}\frac{1}{16}\left(b_0^\text{in}-b_0^\text{out}\right)^2\frac{(\pi \tau\lambda)^2}{\sinh^2(\pi\tau k)}
\left\{1+ \frac{\lambda}{k}\real f(k\tau)\right\}\,.
\gal
The first term in the braces is precisely the leading perturbative result \eq{p30}, while the second (NLO) term explicitly depends on the photon polarization $\lambda$. The polarization is a significant effect only if $b_0k\gtrsim 1$ and $b_0\tau\gtrsim1$. This conclusion is borne out by Figs.~\ref{fig0} and \ref{fig1}.

Fig.~\ref{fig0} shows a typical photon spectrum. A notable feature of these spectra is the large disparity between the two polarizations, indicating that the electromagnetic radiation emitted by a chiral medium is polarized. The peaks correspond to photon momenta given by \eq{d31} and \eq{d33}. The helicity $\lambda$ of the resonances is determined by the signs of parameters $b_0^\text{in}$ and $b_0^\text{out}$. Fig.~\ref{fig1} (left) depicts the photon spectra emitted from quark-gluon plasma, which is comparable to other sources of electromagnetic radiation from that system \cite{PHENIX:2022rsx}.  Fig.~\ref{fig1} (right) exhibits the electromagnetic energy density of radiation produced from a typical chiral material, with the value of $b_0$ varying from zero to 8 meV and the pulse duration, appearing in \eq{p25}, set to be  $\tau= 0.66$~ps. The areas under the blue and green lines represent  the total energy density of the right-hand and left-hand polarized components of the radiation. 

\section{Summary and discussion}\label{sec:s}

We computed the electromagnetic radiation emitted by a chiral medium with the time-dependent chiral chemical potential $\mu_5(t)$ given by \eq{p25} with $b_0=c_A\mu_5$. The spectrum, given by \eq{d25}, explicitly depends on the photon polarization. If the chiral chemical potential does not change sign during the time-evolution, the radiation has nearly perfect right-hand polarization when $\mu_5$ is positive, and left-hand polarization when  $\mu_5$ is negative. The radiation spectrum exhibits a resonance structure, reflecting the instability of the electromagnetic field in the chiral medium at finite $\mu_5$. 

Clearly, the instability plays an important role in inducing the polarization of the radiation. However, it is unclear whether the instability is the only source of polarization. Indeed, the perturbative formula \eq{d38} has no divergencies yet predicts the overall polarization of the radiation at the next-to-leading order. 

To assess the phenomenological relevance of polarized radiation by chiral systems, we performed numerical calculations for two distinct systems: the quark-gluon plasma and a typical chiral material. In the former system, the time-evolution of $\mu_5$ is induced by the parallel chromoelectric and chromomagnetic fields during  the initial stage following a heavy-ion collision \cite{Lappi:2017skr} or by a sphaleron transition at a later stage \cite{Fukushima:2008xe}. In the latter system, it can be driven by an external field. We found that the intensity of the produced photons is comparable to that of other sources of direct photons in heavy-ion collisions \cite{PHENIX:2022rsx}.  The polarized electromagnetic radiation produced by chiral materials within the THz frequency range makes it useful in practical applications. 

Another interesting model to consider is $b_0=A\cos (t/\tau)$ leading to the Mathieu equation for the amplitude $a(t)$. This model describes the chiral chemical potential, which is driven by, for instance, a laser field. We defer further exploration of this model and the attendant parametric resonances to future studies.

In the case of Weyl semimetals characterized by finite $\b \Delta(t)$ and $\mu_5=0$, we argued using perturbation theory that  no radiation is emitted at the leading order in $\b \Delta$. Unfortunately, we were unable to apply the method of Bogolyubov transformation in this case, leaving us uncertain whether any radiation can be emitted from Weyl semimetals. It is also noteworthy that since $\mu_5=0$, the electromagnetic field is perfectly stable. Therefore, the question of whether time-varying $\b \Delta(t)$ in Weyl semimetals produces polarized electromagnetic radiation is closely related to the question of whether the chiral instability is the only source of the radiation, as previously discussed.

\acknowledgments
This work  was supported in part by the U.S. Department of Energy under Grant No.\ DE-SC0023692.

\bibliography{anom-biblio}

\begin{thebibliography}{28}%
\makeatletter
\providecommand \@ifxundefined [1]{%
 \@ifx{#1\undefined}
}%
\providecommand \@ifnum [1]{%
 \ifnum #1\expandafter \@firstoftwo
 \else \expandafter \@secondoftwo
 \fi
}%
\providecommand \@ifx [1]{%
 \ifx #1\expandafter \@firstoftwo
 \else \expandafter \@secondoftwo
 \fi
}%
\providecommand \natexlab [1]{#1}%
\providecommand \enquote  [1]{``#1''}%
\providecommand \bibnamefont  [1]{#1}%
\providecommand \bibfnamefont [1]{#1}%
\providecommand \citenamefont [1]{#1}%
\providecommand \href@noop [0]{\@secondoftwo}%
\providecommand \href [0]{\begingroup \@sanitize@url \@href}%
\providecommand \@href[1]{\@@startlink{#1}\@@href}%
\providecommand \@@href[1]{\endgroup#1\@@endlink}%
\providecommand \@sanitize@url [0]{\catcode `\\12\catcode `\$12\catcode
  `\&12\catcode `\#12\catcode `\^12\catcode `\_12\catcode `\%12\relax}%
\providecommand \@@startlink[1]{}%
\providecommand \@@endlink[0]{}%
\providecommand \url  [0]{\begingroup\@sanitize@url \@url }%
\providecommand \@url [1]{\endgroup\@href {#1}{\urlprefix }}%
\providecommand \urlprefix  [0]{URL }%
\providecommand \Eprint [0]{\href }%
\providecommand \doibase [0]{https://doi.org/}%
\providecommand \selectlanguage [0]{\@gobble}%
\providecommand \bibinfo  [0]{\@secondoftwo}%
\providecommand \bibfield  [0]{\@secondoftwo}%
\providecommand \translation [1]{[#1]}%
\providecommand \BibitemOpen [0]{}%
\providecommand \bibitemStop [0]{}%
\providecommand \bibitemNoStop [0]{.\EOS\space}%
\providecommand \EOS [0]{\spacefactor3000\relax}%
\providecommand \BibitemShut  [1]{\csname bibitem#1\endcsname}%
\let\auto@bib@innerbib\@empty
\bibitem [{\citenamefont {Adler}(1969)}]{Adler:1969gk}%
  \BibitemOpen
  \bibfield  {author} {\bibinfo {author} {\bibfnamefont {S.~L.}\ \bibnamefont
  {Adler}},\ }\bibfield  {title} {\bibinfo {title} {{Axial vector vertex in
  spinor electrodynamics}},\ }\href {https://doi.org/10.1103/PhysRev.177.2426}
  {\bibfield  {journal} {\bibinfo  {journal} {Phys. Rev.}\ }\textbf {\bibinfo
  {volume} {177}},\ \bibinfo {pages} {2426} (\bibinfo {year}
  {1969})}\BibitemShut {NoStop}%
\bibitem [{\citenamefont {Bell}\ and\ \citenamefont
  {Jackiw}(1969)}]{Bell:1969ts}%
  \BibitemOpen
  \bibfield  {author} {\bibinfo {author} {\bibfnamefont {J.~S.}\ \bibnamefont
  {Bell}}\ and\ \bibinfo {author} {\bibfnamefont {R.}~\bibnamefont {Jackiw}},\
  }\bibfield  {title} {\bibinfo {title} {{A PCAC puzzle: $\pi^0 \to \gamma
  \gamma$ in the $\sigma$ model}},\ }\href {https://doi.org/10.1007/BF02823296}
  {\bibfield  {journal} {\bibinfo  {journal} {Nuovo Cim. A}\ }\textbf {\bibinfo
  {volume} {60}},\ \bibinfo {pages} {47} (\bibinfo {year} {1969})}\BibitemShut
  {NoStop}%
\bibitem [{\citenamefont {Zubkov}(2024)}]{Zubkov:2023dpe}%
  \BibitemOpen
  \bibfield  {author} {\bibinfo {author} {\bibfnamefont {M.~A.}\ \bibnamefont
  {Zubkov}},\ }\bibfield  {title} {\bibinfo {title} {{Weyl orbits as probe of
  chiral separation effect in magnetic Weyl semimetals}},\ }\href
  {https://doi.org/10.1088/1361-648X/ad5d36} {\bibfield  {journal} {\bibinfo
  {journal} {J. Phys. Condens. Matter}\ }\textbf {\bibinfo {volume} {36}},\
  \bibinfo {pages} {415501} (\bibinfo {year} {2024})},\ \Eprint
  {https://arxiv.org/abs/2311.12712} {arXiv:2311.12712 [cond-mat.mes-hall]}
  \BibitemShut {NoStop}%
\bibitem [{\citenamefont {Fukushima}\ and\ \citenamefont
  {Mameda}(2012)}]{Fukushima:2012fg}%
  \BibitemOpen
  \bibfield  {author} {\bibinfo {author} {\bibfnamefont {K.}~\bibnamefont
  {Fukushima}}\ and\ \bibinfo {author} {\bibfnamefont {K.}~\bibnamefont
  {Mameda}},\ }\bibfield  {title} {\bibinfo {title} {{Wess-Zumino-Witten action
  and photons from the Chiral Magnetic Effect}},\ }\href
  {https://doi.org/10.1103/PhysRevD.86.071501} {\bibfield  {journal} {\bibinfo
  {journal} {Phys. Rev. D}\ }\textbf {\bibinfo {volume} {86}},\ \bibinfo
  {pages} {071501} (\bibinfo {year} {2012})},\ \Eprint
  {https://arxiv.org/abs/1206.3128} {arXiv:1206.3128 [hep-ph]} \BibitemShut
  {NoStop}%
\bibitem [{\citenamefont {Yee}(2013)}]{Yee:2013qma}%
  \BibitemOpen
  \bibfield  {author} {\bibinfo {author} {\bibfnamefont {H.-U.}\ \bibnamefont
  {Yee}},\ }\bibfield  {title} {\bibinfo {title} {{Flows and polarization of
  early photons with magnetic field at strong coupling}},\ }\href
  {https://doi.org/10.1103/PhysRevD.88.026001} {\bibfield  {journal} {\bibinfo
  {journal} {Phys. Rev. D}\ }\textbf {\bibinfo {volume} {88}},\ \bibinfo
  {pages} {026001} (\bibinfo {year} {2013})},\ \Eprint
  {https://arxiv.org/abs/1303.3571} {arXiv:1303.3571 [nucl-th]} \BibitemShut
  {NoStop}%
\bibitem [{\citenamefont {Jia}\ \emph {et~al.}(2023)\citenamefont {Jia},
  \citenamefont {Li},\ and\ \citenamefont {Hou}}]{Jia:2022awu}%
  \BibitemOpen
  \bibfield  {author} {\bibinfo {author} {\bibfnamefont {M.}~\bibnamefont
  {Jia}}, \bibinfo {author} {\bibfnamefont {H.}~\bibnamefont {Li}},\ and\
  \bibinfo {author} {\bibfnamefont {D.}~\bibnamefont {Hou}},\ }\bibfield
  {title} {\bibinfo {title} {{The photon production and collective flows from
  magnetic induced gluon fusion and splitting in early stage of high energy
  nuclear collision}},\ }\href {https://doi.org/10.1016/j.physletb.2023.138239}
  {\bibfield  {journal} {\bibinfo  {journal} {Phys. Lett. B}\ }\textbf
  {\bibinfo {volume} {846}},\ \bibinfo {pages} {138239} (\bibinfo {year}
  {2023})},\ \Eprint {https://arxiv.org/abs/2211.16770} {arXiv:2211.16770
  [hep-ph]} \BibitemShut {NoStop}%
\bibitem [{\citenamefont {Jia}(2025)}]{Jia:2024ctx}%
  \BibitemOpen
  \bibfield  {author} {\bibinfo {author} {\bibfnamefont {M.}~\bibnamefont
  {Jia}},\ }\bibfield  {title} {\bibinfo {title} {{Photon production by chiral
  magnetic effect in the early stage of high-energy nuclear collisions}},\
  }\href {https://doi.org/10.1103/PhysRevD.111.076015} {\bibfield  {journal}
  {\bibinfo  {journal} {Phys. Rev. D}\ }\textbf {\bibinfo {volume} {111}},\
  \bibinfo {pages} {076015} (\bibinfo {year} {2025})},\ \Eprint
  {https://arxiv.org/abs/2412.17567} {arXiv:2412.17567 [hep-ph]} \BibitemShut
  {NoStop}%
\bibitem [{\citenamefont {Wang}\ and\ \citenamefont
  {Shovkovy}(2024)}]{Wang:2024gnh}%
  \BibitemOpen
  \bibfield  {author} {\bibinfo {author} {\bibfnamefont {X.}~\bibnamefont
  {Wang}}\ and\ \bibinfo {author} {\bibfnamefont {I.~A.}\ \bibnamefont
  {Shovkovy}},\ }\bibfield  {title} {\bibinfo {title} {{Circularly polarized
  photon emission from magnetized chiral plasmas}},\ }\href
  {https://doi.org/10.1103/PhysRevD.110.116005} {\bibfield  {journal} {\bibinfo
   {journal} {Phys. Rev. D}\ }\textbf {\bibinfo {volume} {110}},\ \bibinfo
  {pages} {116005} (\bibinfo {year} {2024})},\ \Eprint
  {https://arxiv.org/abs/2407.06271} {arXiv:2407.06271 [hep-ph]} \BibitemShut
  {NoStop}%
\bibitem [{\citenamefont {Kharzeev}\ and\ \citenamefont
  {Loshaj}(2014)}]{Kharzeev:2013wra}%
  \BibitemOpen
  \bibfield  {author} {\bibinfo {author} {\bibfnamefont {D.~E.}\ \bibnamefont
  {Kharzeev}}\ and\ \bibinfo {author} {\bibfnamefont {F.}~\bibnamefont
  {Loshaj}},\ }\bibfield  {title} {\bibinfo {title} {{Anomalous soft photon
  production from the induced currents in Dirac sea}},\ }\href
  {https://doi.org/10.1103/PhysRevD.89.074053} {\bibfield  {journal} {\bibinfo
  {journal} {Phys. Rev. D}\ }\textbf {\bibinfo {volume} {89}},\ \bibinfo
  {pages} {074053} (\bibinfo {year} {2014})},\ \Eprint
  {https://arxiv.org/abs/1308.2716} {arXiv:1308.2716 [hep-ph]} \BibitemShut
  {NoStop}%
\bibitem [{\citenamefont {Tuchin}(2019)}]{Tuchin:2019jxd}%
  \BibitemOpen
  \bibfield  {author} {\bibinfo {author} {\bibfnamefont {K.}~\bibnamefont
  {Tuchin}},\ }\bibfield  {title} {\bibinfo {title} {{Photon radiation in hot
  nuclear matter by means of chiral anomalies}},\ }\href
  {https://doi.org/10.1103/PhysRevC.99.064907} {\bibfield  {journal} {\bibinfo
  {journal} {Phys. Rev. C}\ }\textbf {\bibinfo {volume} {99}},\ \bibinfo
  {pages} {064907} (\bibinfo {year} {2019})},\ \Eprint
  {https://arxiv.org/abs/1903.02629} {arXiv:1903.02629 [hep-ph]} \BibitemShut
  {NoStop}%
\bibitem [{\citenamefont {Hansen}\ \emph {et~al.}(2025)\citenamefont {Hansen},
  \citenamefont {Ikeda}, \citenamefont {Kharzeev}, \citenamefont {Li},\ and\
  \citenamefont {Tuchin}}]{Hansen:2024kvc}%
  \BibitemOpen
  \bibfield  {author} {\bibinfo {author} {\bibfnamefont {J.}~\bibnamefont
  {Hansen}}, \bibinfo {author} {\bibfnamefont {K.}~\bibnamefont {Ikeda}},
  \bibinfo {author} {\bibfnamefont {D.~E.}\ \bibnamefont {Kharzeev}}, \bibinfo
  {author} {\bibfnamefont {Q.}~\bibnamefont {Li}},\ and\ \bibinfo {author}
  {\bibfnamefont {K.}~\bibnamefont {Tuchin}},\ }\bibfield  {title} {\bibinfo
  {title} {{Magnetic Weyl semimetals as a source of circularly polarized THz
  radiation}},\ }\href {https://doi.org/10.1016/j.physo.2025.100268} {\bibfield
   {journal} {\bibinfo  {journal} {Phys. Open}\ }\textbf {\bibinfo {volume}
  {23}},\ \bibinfo {pages} {100268} (\bibinfo {year} {2025})},\ \Eprint
  {https://arxiv.org/abs/2405.11076} {arXiv:2405.11076 [cond-mat.mtrl-sci]}
  \BibitemShut {NoStop}%
\bibitem [{\citenamefont {Hansen}\ and\ \citenamefont
  {Tuchin}(2025)}]{Hansen:2025gzt}%
  \BibitemOpen
  \bibfield  {author} {\bibinfo {author} {\bibfnamefont {J.}~\bibnamefont
  {Hansen}}\ and\ \bibinfo {author} {\bibfnamefont {K.}~\bibnamefont
  {Tuchin}},\ }\bibfield  {title} {\bibinfo {title} {{Time evolution of
  parity-odd cascades in homogeneous Abelian and non-Abelian media with chiral
  imbalance}},\ }\href {https://doi.org/10.1103/1p9s-jtkw} {\bibfield
  {journal} {\bibinfo  {journal} {Phys. Rev. D}\ }\textbf {\bibinfo {volume}
  {112}},\ \bibinfo {pages} {014010} (\bibinfo {year} {2025})},\ \Eprint
  {https://arxiv.org/abs/2503.00933} {arXiv:2503.00933 [hep-ph]} \BibitemShut
  {NoStop}%
\bibitem [{\citenamefont {Wilczek}(1987)}]{Wilczek:1987mv}%
  \BibitemOpen
  \bibfield  {author} {\bibinfo {author} {\bibfnamefont {F.}~\bibnamefont
  {Wilczek}},\ }\bibfield  {title} {\bibinfo {title} {{Two Applications of
  Axion Electrodynamics}},\ }\href
  {https://doi.org/10.1103/PhysRevLett.58.1799} {\bibfield  {journal} {\bibinfo
   {journal} {Phys. Rev. Lett.}\ }\textbf {\bibinfo {volume} {58}},\ \bibinfo
  {pages} {1799} (\bibinfo {year} {1987})}\BibitemShut {NoStop}%
\bibitem [{\citenamefont {Carroll}\ \emph {et~al.}(1990)\citenamefont
  {Carroll}, \citenamefont {Field},\ and\ \citenamefont
  {Jackiw}}]{Carroll:1989vb}%
  \BibitemOpen
  \bibfield  {author} {\bibinfo {author} {\bibfnamefont {S.~M.}\ \bibnamefont
  {Carroll}}, \bibinfo {author} {\bibfnamefont {G.~B.}\ \bibnamefont {Field}},\
  and\ \bibinfo {author} {\bibfnamefont {R.}~\bibnamefont {Jackiw}},\
  }\bibfield  {title} {\bibinfo {title} {{Limits on a Lorentz and Parity
  Violating Modification of Electrodynamics}},\ }\href
  {https://doi.org/10.1103/PhysRevD.41.1231} {\bibfield  {journal} {\bibinfo
  {journal} {Phys. Rev. D}\ }\textbf {\bibinfo {volume} {41}},\ \bibinfo
  {pages} {1231} (\bibinfo {year} {1990})}\BibitemShut {NoStop}%
\bibitem [{\citenamefont {Sikivie}(1984)}]{Sikivie:1984yz}%
  \BibitemOpen
  \bibfield  {author} {\bibinfo {author} {\bibfnamefont {P.}~\bibnamefont
  {Sikivie}},\ }\bibfield  {title} {\bibinfo {title} {{On the Interaction of
  Magnetic Monopoles With Axionic Domain Walls}},\ }\href
  {https://doi.org/10.1016/0370-2693(84)91731-3} {\bibfield  {journal}
  {\bibinfo  {journal} {Phys. Lett. B}\ }\textbf {\bibinfo {volume} {137}},\
  \bibinfo {pages} {353} (\bibinfo {year} {1984})}\BibitemShut {NoStop}%
\bibitem [{\citenamefont {Tuchin}(2025)}]{Tuchin:2025lbz}%
  \BibitemOpen
  \bibfield  {author} {\bibinfo {author} {\bibfnamefont {K.}~\bibnamefont
  {Tuchin}},\ }\bibfield  {title} {\bibinfo {title} {{Chiral Cherenkov
  radiation at time-dependent chiral chemical potential}},\ }\href@noop {} {\
  (\bibinfo {year} {2025})},\ \Eprint {https://arxiv.org/abs/2507.07324}
  {arXiv:2507.07324 [hep-ph]} \BibitemShut {NoStop}%
\bibitem [{\citenamefont {Bernard}\ and\ \citenamefont
  {Duncan}(1977)}]{Bernard:1977reg}%
  \BibitemOpen
  \bibfield  {author} {\bibinfo {author} {\bibfnamefont {C.}~\bibnamefont
  {Bernard}}\ and\ \bibinfo {author} {\bibfnamefont {A.}~\bibnamefont
  {Duncan}},\ }\bibfield  {title} {\bibinfo {title} {Regularization and
  renormalization of quantum field theory in curved space-time},\ }\href@noop
  {} {\bibfield  {journal} {\bibinfo  {journal} {Annals of physics}\ }\textbf
  {\bibinfo {volume} {107}},\ \bibinfo {pages} {201} (\bibinfo {year}
  {1977})}\BibitemShut {NoStop}%
\bibitem [{\citenamefont {Joyce}\ and\ \citenamefont
  {Shaposhnikov}(1997)}]{Joyce:1997uy}%
  \BibitemOpen
  \bibfield  {author} {\bibinfo {author} {\bibfnamefont {M.}~\bibnamefont
  {Joyce}}\ and\ \bibinfo {author} {\bibfnamefont {M.~E.}\ \bibnamefont
  {Shaposhnikov}},\ }\bibfield  {title} {\bibinfo {title} {{Primordial magnetic
  fields, right-handed electrons, and the Abelian anomaly}},\ }\href
  {https://doi.org/10.1103/PhysRevLett.79.1193} {\bibfield  {journal} {\bibinfo
   {journal} {Phys. Rev. Lett.}\ }\textbf {\bibinfo {volume} {79}},\ \bibinfo
  {pages} {1193} (\bibinfo {year} {1997})},\ \Eprint
  {https://arxiv.org/abs/astro-ph/9703005} {arXiv:astro-ph/9703005}
  \BibitemShut {NoStop}%
\bibitem [{\citenamefont {Boyarsky}\ \emph {et~al.}(2012)\citenamefont
  {Boyarsky}, \citenamefont {Frohlich},\ and\ \citenamefont
  {Ruchayskiy}}]{Boyarsky:2011uy}%
  \BibitemOpen
  \bibfield  {author} {\bibinfo {author} {\bibfnamefont {A.}~\bibnamefont
  {Boyarsky}}, \bibinfo {author} {\bibfnamefont {J.}~\bibnamefont {Frohlich}},\
  and\ \bibinfo {author} {\bibfnamefont {O.}~\bibnamefont {Ruchayskiy}},\
  }\bibfield  {title} {\bibinfo {title} {{Self-consistent evolution of magnetic
  fields and chiral asymmetry in the early Universe}},\ }\href
  {https://doi.org/10.1103/PhysRevLett.108.031301} {\bibfield  {journal}
  {\bibinfo  {journal} {Phys. Rev. Lett.}\ }\textbf {\bibinfo {volume} {108}},\
  \bibinfo {pages} {031301} (\bibinfo {year} {2012})},\ \Eprint
  {https://arxiv.org/abs/1109.3350} {arXiv:1109.3350 [astro-ph.CO]}
  \BibitemShut {NoStop}%
\bibitem [{\citenamefont {Hirono}\ \emph {et~al.}(2015)\citenamefont {Hirono},
  \citenamefont {Kharzeev},\ and\ \citenamefont {Yin}}]{Hirono:2015rla}%
  \BibitemOpen
  \bibfield  {author} {\bibinfo {author} {\bibfnamefont {Y.}~\bibnamefont
  {Hirono}}, \bibinfo {author} {\bibfnamefont {D.}~\bibnamefont {Kharzeev}},\
  and\ \bibinfo {author} {\bibfnamefont {Y.}~\bibnamefont {Yin}},\ }\bibfield
  {title} {\bibinfo {title} {{Self-similar inverse cascade of magnetic helicity
  driven by the chiral anomaly}},\ }\href
  {https://doi.org/10.1103/PhysRevD.92.125031} {\bibfield  {journal} {\bibinfo
  {journal} {Phys. Rev. D}\ }\textbf {\bibinfo {volume} {92}},\ \bibinfo
  {pages} {125031} (\bibinfo {year} {2015})},\ \Eprint
  {https://arxiv.org/abs/1509.07790} {arXiv:1509.07790 [hep-th]} \BibitemShut
  {NoStop}%
\bibitem [{\citenamefont {Akamatsu}\ and\ \citenamefont
  {Yamamoto}(2013)}]{Akamatsu:2013pjd}%
  \BibitemOpen
  \bibfield  {author} {\bibinfo {author} {\bibfnamefont {Y.}~\bibnamefont
  {Akamatsu}}\ and\ \bibinfo {author} {\bibfnamefont {N.}~\bibnamefont
  {Yamamoto}},\ }\bibfield  {title} {\bibinfo {title} {{Chiral Plasma
  Instabilities}},\ }\href {https://doi.org/10.1103/PhysRevLett.111.052002}
  {\bibfield  {journal} {\bibinfo  {journal} {Phys. Rev. Lett.}\ }\textbf
  {\bibinfo {volume} {111}},\ \bibinfo {pages} {052002} (\bibinfo {year}
  {2013})},\ \Eprint {https://arxiv.org/abs/1302.2125} {arXiv:1302.2125
  [nucl-th]} \BibitemShut {NoStop}%
\bibitem [{\citenamefont {Tuchin}(2015)}]{Tuchin:2014iua}%
  \BibitemOpen
  \bibfield  {author} {\bibinfo {author} {\bibfnamefont {K.}~\bibnamefont
  {Tuchin}},\ }\bibfield  {title} {\bibinfo {title} {{Electromagnetic field and
  the chiral magnetic effect in the quark-gluon plasma}},\ }\href
  {https://doi.org/10.1103/PhysRevC.91.064902} {\bibfield  {journal} {\bibinfo
  {journal} {Phys. Rev. C}\ }\textbf {\bibinfo {volume} {91}},\ \bibinfo
  {pages} {064902} (\bibinfo {year} {2015})},\ \Eprint
  {https://arxiv.org/abs/1411.1363} {arXiv:1411.1363 [hep-ph]} \BibitemShut
  {NoStop}%
\bibitem [{\citenamefont {Buividovich}\ and\ \citenamefont
  {Ulybyshev}(2016)}]{Buividovich:2015jfa}%
  \BibitemOpen
  \bibfield  {author} {\bibinfo {author} {\bibfnamefont {P.~V.}\ \bibnamefont
  {Buividovich}}\ and\ \bibinfo {author} {\bibfnamefont {M.~V.}\ \bibnamefont
  {Ulybyshev}},\ }\bibfield  {title} {\bibinfo {title} {{Numerical study of
  chiral plasma instability within the classical statistical field theory
  approach}},\ }\href {https://doi.org/10.1103/PhysRevD.94.025009} {\bibfield
  {journal} {\bibinfo  {journal} {Phys. Rev. D}\ }\textbf {\bibinfo {volume}
  {94}},\ \bibinfo {pages} {025009} (\bibinfo {year} {2016})},\ \Eprint
  {https://arxiv.org/abs/1509.02076} {arXiv:1509.02076 [hep-th]} \BibitemShut
  {NoStop}%
\bibitem [{\citenamefont {Manuel}\ and\ \citenamefont
  {Torres-Rincon}(2015)}]{Manuel:2015zpa}%
  \BibitemOpen
  \bibfield  {author} {\bibinfo {author} {\bibfnamefont {C.}~\bibnamefont
  {Manuel}}\ and\ \bibinfo {author} {\bibfnamefont {J.~M.}\ \bibnamefont
  {Torres-Rincon}},\ }\bibfield  {title} {\bibinfo {title} {{Dynamical
  evolution of the chiral magnetic effect: Applications to the quark-gluon
  plasma}},\ }\href {https://doi.org/10.1103/PhysRevD.92.074018} {\bibfield
  {journal} {\bibinfo  {journal} {Phys. Rev. D}\ }\textbf {\bibinfo {volume}
  {92}},\ \bibinfo {pages} {074018} (\bibinfo {year} {2015})},\ \Eprint
  {https://arxiv.org/abs/1501.07608} {arXiv:1501.07608 [hep-ph]} \BibitemShut
  {NoStop}%
\bibitem [{\citenamefont {Kirilin}\ and\ \citenamefont
  {Sadofyev}(2017)}]{Kirilin:2017tdh}%
  \BibitemOpen
  \bibfield  {author} {\bibinfo {author} {\bibfnamefont {V.~P.}\ \bibnamefont
  {Kirilin}}\ and\ \bibinfo {author} {\bibfnamefont {A.~V.}\ \bibnamefont
  {Sadofyev}},\ }\bibfield  {title} {\bibinfo {title} {{Anomalous Transport and
  Generalized Axial Charge}},\ }\href
  {https://doi.org/10.1103/PhysRevD.96.016019} {\bibfield  {journal} {\bibinfo
  {journal} {Phys. Rev. D}\ }\textbf {\bibinfo {volume} {96}},\ \bibinfo
  {pages} {016019} (\bibinfo {year} {2017})},\ \Eprint
  {https://arxiv.org/abs/1703.02483} {arXiv:1703.02483 [hep-th]} \BibitemShut
  {NoStop}%
\bibitem [{\citenamefont {Abdulameer}\ \emph {et~al.}(2024)\citenamefont
  {Abdulameer} \emph {et~al.}}]{PHENIX:2022rsx}%
  \BibitemOpen
  \bibfield  {author} {\bibinfo {author} {\bibfnamefont {N.~J.}\ \bibnamefont
  {Abdulameer}} \emph {et~al.} (\bibinfo {collaboration} {PHENIX}),\ }\bibfield
   {title} {\bibinfo {title} {{Nonprompt direct-photon production in Au+Au
  collisions at sNN=200 GeV}},\ }\href
  {https://doi.org/10.1103/PhysRevC.109.044912} {\bibfield  {journal} {\bibinfo
   {journal} {Phys. Rev. C}\ }\textbf {\bibinfo {volume} {109}},\ \bibinfo
  {pages} {044912} (\bibinfo {year} {2024})},\ \Eprint
  {https://arxiv.org/abs/2203.17187} {arXiv:2203.17187 [nucl-ex]} \BibitemShut
  {NoStop}%
\bibitem [{\citenamefont {Lappi}\ and\ \citenamefont
  {Schlichting}(2018)}]{Lappi:2017skr}%
  \BibitemOpen
  \bibfield  {author} {\bibinfo {author} {\bibfnamefont {T.}~\bibnamefont
  {Lappi}}\ and\ \bibinfo {author} {\bibfnamefont {S.}~\bibnamefont
  {Schlichting}},\ }\bibfield  {title} {\bibinfo {title} {{Linearly polarized
  gluons and axial charge fluctuations in the Glasma}},\ }\href
  {https://doi.org/10.1103/PhysRevD.97.034034} {\bibfield  {journal} {\bibinfo
  {journal} {Phys. Rev. D}\ }\textbf {\bibinfo {volume} {97}},\ \bibinfo
  {pages} {034034} (\bibinfo {year} {2018})},\ \Eprint
  {https://arxiv.org/abs/1708.08625} {arXiv:1708.08625 [hep-ph]} \BibitemShut
  {NoStop}%
\bibitem [{\citenamefont {Fukushima}\ \emph {et~al.}(2008)\citenamefont
  {Fukushima}, \citenamefont {Kharzeev},\ and\ \citenamefont
  {Warringa}}]{Fukushima:2008xe}%
  \BibitemOpen
  \bibfield  {author} {\bibinfo {author} {\bibfnamefont {K.}~\bibnamefont
  {Fukushima}}, \bibinfo {author} {\bibfnamefont {D.~E.}\ \bibnamefont
  {Kharzeev}},\ and\ \bibinfo {author} {\bibfnamefont {H.~J.}\ \bibnamefont
  {Warringa}},\ }\bibfield  {title} {\bibinfo {title} {{The Chiral Magnetic
  Effect}},\ }\href {https://doi.org/10.1103/PhysRevD.78.074033} {\bibfield
  {journal} {\bibinfo  {journal} {Phys. Rev. D}\ }\textbf {\bibinfo {volume}
  {78}},\ \bibinfo {pages} {074033} (\bibinfo {year} {2008})},\ \Eprint
  {https://arxiv.org/abs/0808.3382} {arXiv:0808.3382 [hep-ph]} \BibitemShut
  {NoStop}%
\end{thebibliography}%


\end{document}